\def\ra{\rangle}
\def\la{\langle}
\def\be{\begin{equation}}
\def\ee{\end{equation}}
\def\ba{\begin{array}}
\def\ea{\end{array}}

\documentclass[prl,showpacs,preprint,amsmath,amssymb]{revtex4}
\usepackage{graphicx}
\usepackage{caption2}
\def\qed{\leavevmode\unskip\penalty9999 \hbox{}\nobreak\hfill
     \quad\hbox{\leavevmode  \hbox to.77778em{%
               \hfil\vrule   \vbox to.675em%
               {\hrule width.6em\vfil\hrule}\vrule\hfil}}
     \par\vskip3pt}

\begin{document}
\title{Lower Bound of Concurrence Based on Generalized Positive Maps}
\author{Hui-Hui Qin$^{1}$}
\author{Shao-Ming Fei$^{2,3}$}

\affiliation{$^1$Department of Mathematics, School of Science, South
China University of Technology, Guangzhou 510640, China\\
$^2$School of Mathematical Sciences, Capital Normal University,
Beijing 100048, China\\
$^3$Max-Planck-Institute for Mathematics in the Sciences, 04103
Leipzig, Germany}

\begin{abstract}

We study the concurrence of arbitrary dimensional bipartite quantum
systems. By using a positive but not completely
positive map, we present an analytical lower bound of concurrence.
Detailed examples are used to show that our bound can detect
entanglement better and can improve the well known existing lower bounds.

\end{abstract}
\maketitle

Quantum entanglement plays significant roles in quantum
information processing \cite{hhhh}. The
concurrence \cite{con} is one of the important measures
of quantum entanglement. It plays an essential role in describing quantum
phase transitions in various interacting quantum many-body systems
\cite{Osterloh02-Wu04,Ghosh2003}.
However, due to the extremizations involved in the calculation,
for general high dimensional case only a few explicit analytic formulae for
concurrence have been found
for some special symmetric states \cite{Terhal-Voll2000}.

To estimate the concurrence for general bipartite states,
the lower bounds of concurrence have been extensively studied
\cite{167902,Chen-Albeverio-Fei1,chen,breuerprl,breuer,vicente,zhang,edward,ou,Li,zhao-S.M.,Gao,Yan}.
In \cite{Li} a lower bound of concurrence based on a positive map was obtained,
which is better than other lower bounds for some quantum states.
In this paper we use a series of generalized positive maps which include
the one in \cite{Li} as a special case.
We show that these generalized maps can also give rise to lower bounds
of concurrence which improves the existing ones.

Let $H_1$ and $H_2$ be $n$-dimensional vector spaces.
A bipartite quantum pure state $|\phi\ra$ in $H_1\otimes H_2$ has a
Schmidt form
\begin{equation}
\label{schmit}
|\phi\ra=\sum_i \alpha_i|e_i^1\ra \otimes|e_i^2\ra,
\end{equation}
where $|e_i^1\ra$ and  $|e_i^2\ra$  are the orthonormal bases in $H_1$ and
$H_2$ respectively, $\alpha_i$ are the Schmidt coefficients satisfying $\sum_i\alpha_i^2=1$.
The concurrence of the state $|\psi\ra$ is given by
\be\label{concurrence}
C(|\phi\ra)=\sqrt{2(1-Tr\rho_1^2})=2\sqrt{\sum_{i<
j}\alpha_i^2\alpha_j^2},
\ee
where $\rho_1$ is the reduced density matrix
obtained by tracing over the second subsystem of the
density matrix $\rho=|\phi\ra\la \phi|$, $\rho_1=Tr_2|\phi\ra\la\phi|$.

A general mixed state in  $H_1\otimes H_2$ has pure state
decompositions, $\rho=\sum_i p_i|\phi_i\ra \la\phi_i|$, where $p_i
\geq 0$ and $\sum_i\ p_i=1$. The concurrence is extended to mixed
states $\rho$ by the convex roof,
\be\label{concurrencem}
C(\rho)=\min_{\{p_{i},|\phi_{i}\rangle\}}\sum_{i}p_{i}C(|\phi_{i}\rangle).
\ee
where the minimum is taken over all possible pure state
decompositions ${\{p_{i},|\phi_{i}\rangle\}}$ of $\rho$.

Let $f(\rho)$ be a real-valued and convex function of $\rho$ such that for any pure state
$|\phi\ra$ with Schmidt decomposition (\ref{schmit}),
\begin{equation} \label{property}
f(|\phi\ra \la\phi|)\leq 2\sum_{i< j}\alpha_i\alpha_j.
\end{equation}
Breuer derived in \cite{breuer} that $C(\rho)$ satisfies
\be\label{bb}
C(\rho)\geq\sqrt{\frac{2}{N(N-1)}} f(\rho).
\ee

The $f(\rho)$ corresponding to the lower bounds in \cite{chen} are the ones
with respect to the PPT criterion and the realignment criterion,
$f_{ppt}(\rho) = ||\rho^{T_1}||-1$, $f_{r}(\rho) =
||\tilde{\rho}||-1$, where $||\cdot||$ stands for the trace norm of
a matrix, $T_1$ the partial transposition associated with the space
$H_1$ and $\tilde{\rho}$ the realigned matrix of $\rho$. Namely
\begin{equation}\label{eq-A}
C_{PPT}(\rho) \geq \sqrt{\frac{2}{n(n-1)}}(\|\rho^{T_{1}}\|-1),
\end{equation}
\begin{equation}\label{eq-B}
C_{r}(\rho)\geq \sqrt{\frac{2}{n(n-1)}}(\|\widetilde{\rho}\|-1).
\end{equation}
The lower bound obtained in \cite{Li} corresponds
to $f_1(\rho)=\|(I\otimes\Phi)\rho\|-(n-1)$,
\begin{equation}\label{eq-C}
C_{1}(\rho) \geq \sqrt{\frac{2}{n(n-1)}}[\|(I\otimes\Phi)\rho\|-(n-1)],
\end{equation}
where the positive but not completely positive map $\Phi$
maps an $n\times n$ matrix $A$, $(A)_{ij}=a_{ij}$, $i,j=1,...n$,
to an $n\times n$ matrix $\Phi(A)$ with $(\Phi(A))_{ij}=-a_{ij}$ for $i\neq j$, and
$(\Phi(A))_{ii}=(n-2)a_{ii}+a_{i^\prime i^\prime}$, $i^\prime=i+1(~mod~ n)$,
\begin{equation}\label{Phi}
\begin{array}{rcl}
\Phi(A)&=&\displaystyle (n-1)\sum^{n}_{i=1}\emph{E}_{ii}A\emph{E}_{ii}+\sum^{n}_{i=1}\emph{E}_{i,i+1}A\emph{E}_{i,i+1} \\[2mm]
&&\displaystyle -(\sum^{n}_{i=1}\emph{E}_{ii})A(\sum^{n}_{i=1}\emph{E}_{ii}),
\end{array}
\end{equation}
$E_{ij}$ is the matrix with the $(i,j)$ entry $1$ and the other entries $0$.

We consider the linear map $\Phi_{t,\pi}$ defined by
\begin{widetext}
\begin{equation}\label{Phit}
\Phi_{t,\pi}(X)=
\begin{pmatrix}
a_{11}&-x_{12}&\cdots&-x_{1n}\\
-x_{21}&a_{22}&\cdots&-x_{2n}\\
\vdots&\vdots&\ddots&\vdots\\
-x_{n1}&-x_{n2}&\cdots&a_{nn}
\end{pmatrix},
\end{equation}
\end{widetext}
where $X=(x_{ij})\in \mathbf{M}_{n}(\mathbf{C})$ is any $n\times n$ complex matrix,
$a_{ii}=(n-1-t)x_{ii}+tx_{\pi(i),\pi(i)}$, $i=1,...,n$,
$0\leq t \leq n$ and $\pi$ is any permutation of $(1,2,\ldots,n)$.
When $t=1$, the map $\Phi_{t,\pi}$ is reduced to $\Phi$ in (\ref{Phi}).

According to \cite{houjinchuan}, $\Phi_{t,\pi}$ is positive if and only if $0 \leq t
\leq \frac{n}{l(\pi)}$, where $l(\pi)$ is the length of $\pi$. $\pi$ is said to be cyclic if
$l(\pi)=n$. It has been shown that the map corresponds to the optimal witness when
$l(\pi)=n$ for $n=3$ \cite{houjinchuan}. In the following we consider the case that
$\pi$ is cyclic, i.e. $0\leq t \leq 1$. Without loss of generality,
we assume that the cyclic $\pi$ is defined by $\pi(i)=i+1 (mod n)$, $i=1,2,\ldots,n$.

{\bf Theorem}.  For any bipartite quantum state
$\rho\in\mathcal{}{H_{1}}\otimes\mathcal{}{H_{2}}$, the concurrence
$C(\rho)$ satisfies
\begin{equation}\label{eq-D}
C(\rho)\geq \sqrt{\frac{2}{n(n-1)}}[\|(I_{n}\otimes\Phi_{t,\pi})\rho\|-(n-1)],
\end{equation}
where $I_{n}$ is the $n\times n$ identity matrix, $\pi$ is cyclic and $0\leq t \leq 1$.

{\sf Proof}.  Set
$f(\rho)=\|(I_{n}\otimes\Phi_{t,\pi})\rho\|-(n-1)$. It is apparent
that $f(\rho)$ is real-valued and convex due to the convexity of the trace norm. What we need is to
show that for any pure state (\ref{schmit}),
the inequality (\ref{property}) holds.

As the trace norm does not change under local coordinate transformations, we can take
$|\phi\rangle=(\alpha_{1}, 0, \cdots, 0;0,\alpha_{2}, 0, \cdots, 0;
0, 0, \alpha_{3}, 0,\cdots,0;\cdots;0,\cdots,0,\alpha_{n})^t$, where
$t$ denotes transpose, the Schmidt coefficients satisfy
$0\leq\alpha_{i}\leq1,(i=1,2,\cdots,n)$ and $\sum^{n}_{i=1}\alpha^{2}_{i}=1$.

It is direct to verify that
$I_{n}\otimes\Phi_{t,\pi}(|\phi\rangle\langle\phi|)$ has $n^2-2n$
eigenvalues $0$ and $n$  eigenvalues $t\alpha^{2}_{1},
t\alpha^{2}_{2}, \cdots, t\alpha^{2}_{n}$. And the rest $n$
eigenvalues are given by the eigenvalues of the following matrix,
\begin{equation}
B=
\begin{pmatrix}
(n-1-t)\alpha^{2}_{1}&-\alpha_{1}\alpha_{2}&\cdots&-\alpha_{1}\alpha_{n}\\
-\alpha_{1}\alpha_{2}&(n-1-t)\alpha^{2}_{2}&\cdots&-\alpha_{2}\alpha_{n}\\
\vdots&\vdots&\ddots&\vdots\\
-\alpha_{n}\alpha_{1}&-\alpha_{n}\alpha_{2}&\cdots&(n-1-t)\alpha^{2}_{n}\\
\end{pmatrix}.\nonumber\\
\end{equation}
The eigenpolynomial equation of $B$ is given by
\begin{widetext}
\begin{equation}
\begin{aligned}
g(\lambda)=&|\lambda I_{n}-B|=\lambda^{n}-(n-1-t)\lambda^{n-1}+(n-t)(n-2-t)(\sum_{i<j}\alpha^{2}_{i}\alpha^{2}_{j})\lambda^{n-2} \\
\quad&+\cdots+(-1)^{k}(n-t)^{k-1}(n-1-k-t)(\sum_{i_{1}<i_{2}<\cdots<i_{k}}\alpha^{2}_{i_{1}}\alpha^{2}_{i_{2}}\cdots\alpha^{2}_{i_{k}})\lambda^{n-k} \\
\quad&+\cdots+ (-1)^{n-1}(n-t)^{n-2}(1-t)(\sum_{i_{1}<i_{2}<\cdots<i_{n-1}}\alpha^{2}_{i_{1}}\alpha^{2}_{i_{2}}
\cdots\alpha^{2}_{i_{n-1}})\lambda\\
\quad&+(-1)^{n+1}t(n-t)^{n-1}\prod^{n}_{i=1}\alpha^{2}_{i}.
\end{aligned}
\end{equation}
\end{widetext}

Let $\lambda_{1},\lambda_{2}, \cdots, \lambda_{n}$, $\lambda_{1}\leq\lambda_{2}\leq\ldots\leq\lambda_{n}$,
be the roots of the equation $g(\lambda)=0$. We have
\begin{equation}\label{eq-E}
\begin{aligned}
\sum^{n}_{i=1}\lambda_{i}&=n-1-t, \\
\prod^{n}_{i=1}\lambda_{i}&=(-1)^{2n+1}t(n-t)^{n-1}\prod^{n}_{i=1}\alpha^{2}_{i}.
\end{aligned}
\end{equation}
The inequality (\ref{property}) we need to prove has the form now,
\begin{equation}\label{eq-F}
\sum^{n}_{i=1}|\lambda_{i}|+t-(n-1)\leq2\sum_{i<j}\alpha_{i}\alpha_{j}.
\end{equation}

Set $\beta=\prod^{n}_{i=1}\alpha^{2}_{i}$.
If $\beta=0$, then $g(0)=0$,  $0$ is an eigenvalue of $B$.
From the derivation of $g(\lambda)$ with respect to $\lambda$, we have
\begin{widetext}
\begin{equation}
\begin{aligned}
g'(\lambda)=&\,\, n\lambda^{n-1}-(n-1)(n-1-t)\lambda^{n-2}+(n-2)(n-t)(n-2-t)(\sum_{i<j}\alpha^{2}_{i}\alpha^{2}_{j})\lambda^{n-3} \\
\quad&+\cdots+(-1)^k(n-k)(n-t)^{k-1}(n-1-t)(\sum_{i_{1}<i_{2}<\cdots<i_{k}}\alpha^{2}_{i_{1}}\alpha^{2}_{i_{2}}\cdots\alpha^{2}_{i_{k}})\lambda^{n-k-1} \\
\quad&+\cdots+(-1)^{n-1}(n-t)^{n-2}(1-t)(\sum_{i_{1}<i_{2}<\cdots<i_{n-1}}\alpha^{2}_{i_{1}}\alpha^{2}_{i_{2}}\cdots\alpha^{2}_{i_{n-1}}).
\end{aligned}
\end{equation}
\end{widetext}

If $n$ is even, for all $\lambda\leq0$, we have $g'(\lambda)\leq0$,
that is $g(\lambda)$ is monotonically decreasing for $\lambda\leq0$.
Taking $g(0)=0$ into account, we obtain that $g(\lambda)=0$ has no
negative root, then the inequality ~(\ref{eq-F}) becomes:
\begin{equation}\label{eq-G}
\sum^{n}_{i=1}\lambda_{i}+t-(n-1)\leq2\sum_{i<j}\alpha_{i}\alpha_{j}.
\end{equation}
According to the equations (\ref{eq-E}), (\ref{eq-G}) always holds.

If $n$ is odd, for all $\lambda\leq0$, we have $g'(\lambda)\geq0$,
which means that $g'(\lambda)$ is monotonically increasing  for
$\lambda\leq0$. Hence $g(\lambda)=0$ has no negative root as well,
and the inequality (\ref{eq-G}) also holds.

If $\beta\neq0$, $g'(\lambda)$ is a monotonic
function when $\lambda\leq0$.  From
$g(0)=(-1)^{n+1}t(n-t)^{n-1}\prod^{n}_{i=1}\alpha^{2}_{i}$, we can
get that the equation $g(\lambda)=0$ has only one negative
root $\lambda_{1}$. The inequality (\ref{eq-F}) becomes
\begin{equation}
\sum^{n}_{i=1}\lambda_{i}-2\lambda_{1}+t-(n-1)\leq2\sum_{i<j}\alpha_{i}\alpha_{j}.
\end{equation}
To prove the above inequality, we only need to prove
$\lambda_{1}\geq-\sum_{i<j}\alpha_{i}\alpha_{j}$ by using of
the equations (\ref{eq-E}). From the definition of the $g(\lambda)$,
we have
$g(-\sum_{i<j}\alpha_{i}\alpha_{j})=|-\sum_{i<j}\alpha_{i}\alpha_{j}I_{n}-B|=(-1)^{n}|\sum_{i<j}\alpha_{i}\alpha_{j}I_{n}+B|$.
Due to the property of the diagonally dominant matrix
$\sum_{i<j}\alpha_{i}\alpha_{j}I_{n}+B$,
$|\sum_{i<j}\alpha_{i}\alpha_{j}I_{n}+B|\geq0$ when $n$ is even.
We can get that
$\lambda_{1}\geq-\sum_{i<j}\alpha_{i}\alpha_{j}$ as $g(\lambda)$ is
monotonically decreasing when $\lambda\leq0$. In the same way one can
prove the result when $n$ is odd. \hfill$\Box$

As the positive map $\Phi_{t,\pi}$ in (\ref{Phit}) includes the map $\Phi$ in (\ref{Phi})
as a special case, our lower bound (\ref{eq-D}) is a generalized form of (\ref{eq-C}) in
\cite{Li}. Therefore all states whose entanglement can be identified by
\cite{Li} can be also identified by our lower bound (\ref{eq-D}).
In fact, the lower bound (\ref{eq-D}) can
detect entanglement that other lower bounds cannot. Let us consider
a state of $n=3$,
\begin{equation}
\rho=\frac{1}{(4x+5y)}
\begin{pmatrix}
y&0&0&0&0&0&0&0&0\\
0&x&0&x&0&x&0&x&0\\
0&0&y&0&0&0&0&0&0\\
0&x&0&x&0&x&0&x&0\\
0&0&0&0&y&0&0&0&0\\
0&x&0&x&0&x&0&x&0\\
0&0&0&0&0&0&y&0&0\\
0&x&0&x&0&x&0&x&0\\
0&0&0&0&0&0&0&0&y\\
\end{pmatrix},\nonumber
\end{equation}
where $x>0$ and $y>0$. Under the positive map $\Phi_{t,\pi}$, the density matrix
$\rho'=(I_{3}\otimes\Phi_{t,\pi})\rho$ has the following form,
\begin{widetext}
\begin{equation}\label{rhoe}
\rho'=\frac{1}{(4x+5y)}
\begin{pmatrix}
a&0&0&0&0&0&0&0&0\\
0&b&0&-x&0&-x&0&-x&0\\
0&0&2y&0&0&0&0&0&0\\
0&-x&0&b&0&-x&0&-x&0\\
0&0&0&0&a&0&0&0&0\\
0&-x&0&-x&0&2x&0&-x&0\\
0&0&0&0&0&0&a&0&0\\
0&-x&0&-x&0&-x&0&b&0\\
0&0&0&0&0&0&0&0&2y\\
\end{pmatrix},\nonumber
\end{equation}
\end{widetext}
where $a=(2-t)y+tx$ and $b=(2-t)x+ty$.
The set of eigenvalues of $\rho'$ is given by
\begin{equation}
\begin{aligned}
\displaystyle
\lambda_{1}=&\lambda_{2}=\frac{2y}{4x+5y}, \\
\displaystyle
\lambda_{3}=&\lambda_{4}=\lambda_{5}=\frac{(2-t)y+tx}{4x+5y}, \\
\displaystyle
\lambda_{6}=&\lambda_{7}=\frac{(3-t)x+ty}{4x+5y},\\
\displaystyle
\lambda_{8,9}=& \frac{1}{4x+5y}\Big[(2-t)x+ty \\
 &\pm \sqrt{[(2-t)x+ty]^2+4[(3+2t)x^2-2txy]}\Big]. \nonumber
\end{aligned}
\end{equation}
For $(3+2t)x-2ty>0$, from (\ref{eq-D}) the concurrence of $\rho$ satisfies,
\begin{equation}\label{eq-H}
\begin{aligned}
\displaystyle C(\rho)\geq &\frac{1}{4x+5y}\Big[(2-t)x+ty \\
&\pm \sqrt{[(2-t)x+ty]^2+4[(3+2t)x^2-2txy]}\Big].
\end{aligned}
\end{equation}
From the lower bound of concurrence in \cite{Li} one has,
\begin{equation}\label{eq-I}
C(\rho)\geq \frac{-(x+y)+\sqrt{(x+y)^2+4[5x^2-2xy]}}{4x+5y}.
\end{equation}
The lower bound (\ref{eq-A}) gives rise to
\begin{equation}\label{eq-J}
C(\rho)\geq \frac{2(2x-y)}{4x+5y}.
\end{equation}
While from lower bound (\ref{eq-B}) one has
\begin{equation}\label{eq-K}
C(\rho)\geq \frac{2\sqrt{4x^2+y^2}-4y}{4x+5y}.
\end{equation}

\begin{figure}[htpb]\label{eps1}
\renewcommand{\captionlabeldelim}{.}
\renewcommand{\figurename}{Fig.}
\centering\
\includegraphics[width=8cm]{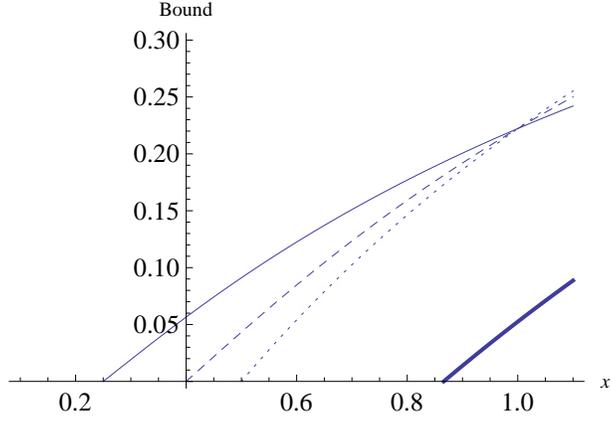}
\caption{{\small The lower bound of concurrence of (\ref{rhoe}), solid line for bound (\ref{eq-H}),
dashed line for bound (\ref{eq-I}), dotted line for bound (\ref{eq-J}),
and thick line for bound (\ref{eq-K}).}}
\end{figure}

\begin{figure}[htpb]\label{eps2}
\renewcommand{\captionlabeldelim}{.}
\renewcommand{\figurename}{Fig.}
\centering\
\includegraphics[width=8cm]{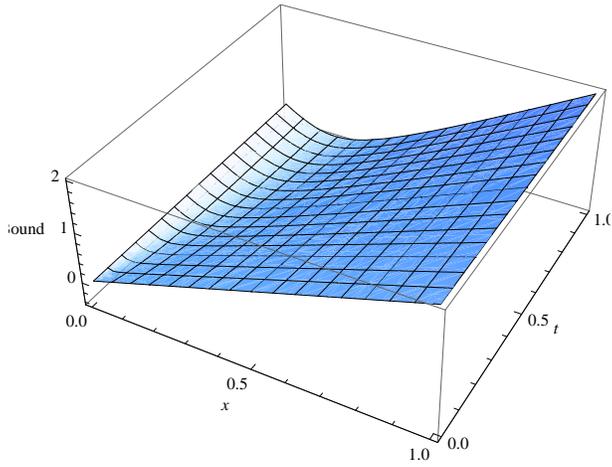}
\caption{{\small The lower bound of concurrence of (\ref{rhoe}) based on the maps $\Phi_{t,\pi}$ for $t\in[0,1]$.}}
\end{figure}

\begin{figure}[htpb]\label{eps3}
\renewcommand{\captionlabeldelim}{.}
\renewcommand{\figurename}{Fig.}
\centering\
\includegraphics[width=8cm]{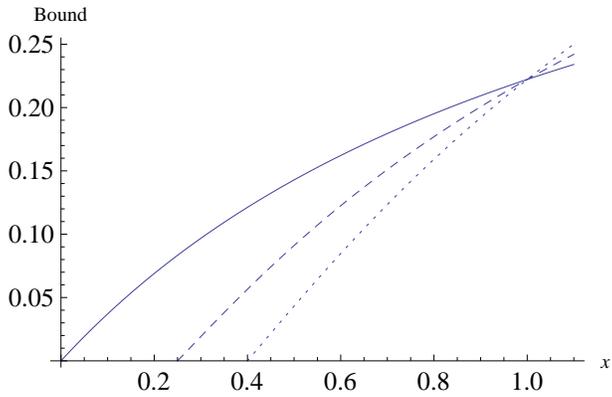}
\caption{{\small The lower bound of concurrence of (\ref{rhoe}) from (\ref{eq-D}).
Solid, dashed and dotted lines correspond respectively to the maps $\Phi_{0,\pi}$, $\Phi_{\frac{1}{2},\pi}$ and $\Phi_{1,\pi}$.}}
\end{figure}

To compare these lower bounds, we take $y=1$ and $t={1}/{2}$.
The lower bounds obtained in \cite{Chen-Albeverio-Fei1,chen,Li} fail to detect the entanglement of
$\rho$ when $\frac{1}{4}< x < \frac{2}{5}$, see Fig.\ref{eps1}.
Our lower bound is better than other lower bounds for $x\in (\frac{1}{4}, 1)$.

The lower bound (\ref{eq-D}) depends on the parameter $t$.
The choice of $t$ depends on detailed quantum states.
Fig.\ref{eps2} shows the entanglement detection ability of (\ref{eq-D})
according to $t$. One can see that when $t=0$ (\ref{eq-D})
can detect the entanglement of (\ref{rhoe}) better, see Fig.\ref{eps3}.

We have presented a new lower bound of concurrence
for arbitrary dimensional bipartite quantum systems, in terms of
a positive but not completely positive map.
The lower bound in \cite{Li} can detect entanglement for some quantum states better
than some well-known separability criteria, and improves the lower
bounds such as from the PPT, realignment criteria and the Breuer's
entanglement witness. Our bound is even better than the one in \cite{Li}, since our bound includes
the bound in \cite{Li} as a special case.
It helps to detect quantum entanglement for certain classes of quantum states.

\bigskip
\noindent{\bf Acknowledgments}\, \, This work is supported by the
NSFC under number 11275131.

\end{document}